\documentclass[a4paper,11pt]{article}
\usepackage{pos}
\usepackage{graphicx}   
\usepackage{multirow}
\usepackage{adjustbox}
\usepackage{array}
\usepackage{booktabs}
\usepackage{bbm}
\usepackage{mathbbol}
\usepackage{enumitem}
\usepackage{mathtools}
\allowdisplaybreaks

\newcolumntype{R}[1]{%
>{\adjustbox{angle=#1}\bgroup}%
l%
<{\egroup}%
}

\renewcommand{\arraystretch}{1.18}

\title{Update on the isospin breaking corrections to the HVP with
$C-$periodic boundary conditions}
\ShortTitle{Update on the IB corrections to the HVP with $C-$periodic boundary conditions}

\author[a]{Anian Altherr}
\author[b]{Isabel Campos}
\author[c,d]{Alessandro Cotellucci}
\author[a]{Roman Gruber}
\author[a]{Tim Harris}
\author[a]{Javad Komijani}
\author[c,e]{Jens L\"ucke}
\author[a]{Marina Krstić Marinković}
\author*[a]{Letizia Parato}
\author[c,e]{Agostino Patella}
\author[b]{Sara Rosso}
\author[a]{Paola Tavella}

\affiliation[a]{Institut für Theoretische Physik, ETH Zürich, Wolfgang-Pauli-Str. 27, 8093 Zürich, Switzerland}
\affiliation[b]{Instituto de Física de Cantabria and IFCA-CSIC, Avda. de Los Castros s/n, 39005 Santander, Spain}
\affiliation[c]{Humboldt Universität zu Berlin, Institut für Physik and IRIS Adlershof, Zum Großen Windkanal 6, 12489
Berlin, Germany}
\affiliation[d]{Jülich Supercomputing Centre, Forschungszentrum Jülich, D-52428 Jülich, Germany}
\affiliation[e]{DESY, Platanenallee 6, D-15738 Zeuthen, Germany}

\emailAdd{lparato@phys.ethz.ch}

\abstract{In the \rcstar collaboration, we simulate lattice QCD+QED using $C-$periodic spatial boundary conditions to ensure that locality, gauge invariance, and translational invariance are preserved throughout the calculation. We present our progress in computing isospin-breaking (IB) corrections to the leading hadronic contribution to $(g-2)_\mu$. We compare two ways of including the IB corrections: the RM123 method and dynamical QCD+QED simulations, both with $C-$periodic boundary conditions. The two calculations are performed at $\beta=3.24$ with four flavours of $\mathcal{O}(a)-$improved Wilson fermions; the QCD ensemble features $SU(3)-$symmetric sea quarks plus charm, while down and strange quarks are degenerate in QCD+QED gauge ensembles. 
}

\FullConference{%
The 41st International Symposium on Lattice Field Theory (LATTICE2024)\\
 28 July - 3 August 2024\\
Liverpool, UK\\}

\newcommand{\rcstar}{RC$^{\star}$~}
\newcommand{\cstar}{C$^{\star}$~}
\newcommand{\hvp}{\text{\scriptsize HVP}}

\begin{document}
\maketitle

\section{Introduction}
\label{sec:intro}

Achieving sub-percent precision in hadronic quantity predictions within lattice QCD requires accounting for isospin-breaking (IB) effects from $m_u \neq m_d$ and from QED, each contributing at the ${\sim1\%}$ level. 
The \rcstar collaboration has developed a framework using $C$-periodic boundary conditions~\cite{Polley:1993bn, Kronfeld:1992ae, Lee:2015rua, Luscher:1985dn} in lattice ensembles, allowing for a lattice formulation of QCD+QED that preserves locality, gauge invariance, and translational invariance. 
Consequently, the collaboration focuses primarily on observables where IB effects are significant, such as the measurement of charged meson masses and the hadronic vacuum polarization (HVP)~\cite{Lucini:2015hfa, Campos:2019kgw, RCstar:2022yjz, RC:2022hxa, RC:2022fly, RC:2023zbm, Dale:2021nxf}.

The goal of this preliminary analysis is to compare two methods for computing observables with IB effects, focusing on their relative performance and uncertainties:
\begin{enumerate}
    \item Monte Carlo sampled, dynamical QCD+QED, where the photon field evolves alongside the SU(3) gauge field in the HMC algorithm. This method requires a setup where $m_u \neq m_d$. 
    \item The RM123 method~\cite{deDivitiis:2012rf, deDivitiis:2013xla}, applied to an ensemble generated in isospin-symmetric QCD (isoQCD), i.e. with $m_u = m_d$ and $e^2 = 0$. 
\end{enumerate}

The observable chosen for this comparison is $a_\mu^\hvp$, the HVP contribution to $(g-2)_\mu$, for which sub-percent precision---and thus inclusion of IB effects---are currently required~\cite{Aoyama:2020ynm}.

The analysis is performed on ensemble \texttt{A380a07b324} ($N_f=1+2+1$) for dynamical QCD+QED and ensemble \texttt{A400a00b324} ($N_f=3+1$) for isoQCD+RM123. Both ensembles, generated in Ref.~\cite{RCstar:2022yjz} using \cstar boundary conditions, share the same simulation setup, except for the inclusion of QED and a slightly lighter $m_u$ in \texttt{A380a07b324}.

We consider two approaches to compare the results obtained from the two methods above:

\begin{enumerate}[label=(\alph*)]
    \item 
    Use theoretical, error-free shifts in the bare parameters $ \vec{\varepsilon} = (\beta, e^2, m_u, m_d, m_s, m_c)$, defined as $\Delta \vec{\varepsilon} = \vec{\varepsilon}_{\text{A380}} - \vec{\varepsilon}_0$,
    where $\vec{\varepsilon}_{\text{A380}}$ and $\vec{\varepsilon}_0$  correspond to the parameters of the QCD+QED (\texttt{A380a07b324}) and isoQCD (\texttt{A400a00b324}) ensembles, respectively. Denoting with $\langle...\rangle_0$ measurements in isoQCD, the shifted quantities
    $\langle \phi_i(\vec \varepsilon_0) \rangle_0 + \Delta \vec{\varepsilon} \cdot \langle \partial_{\vec{\varepsilon}} \phi_i (\vec \varepsilon_0) \rangle_0$ can be compared to those measured on the QCD+QED ensemble, $\langle \phi_i(\vec{\varepsilon}) \rangle$; a similar comparison applies to the gradient flow scale $t_0$. Here, $ \phi_i $ refers to combinations of meson masses, as described in Sec.~\ref{sec:ensembles}. 
    This approach allows us to analyze how the uncertainty is distributed between $a_\mu^\hvp$, the scale-setting parameter $t_0$, and the tuning observables $\phi_i$. Graphically, this comparison corresponds to two points in the $\phi_i$, $t_0$, and $a_\mu^\hvp$ space: one representing the dynamical QCD+QED result and the other the isoQCD+RM123 result, with uncertainties accounted for in all directions.
    \item 
    Compute the shifts $ \Delta \vec{\varepsilon} = \vec{\varepsilon}_* - \vec{\varepsilon}_0 $ (along with their uncertainties) that solve the renormalization conditions
    $ \langle \phi_i(\vec \varepsilon_0) \rangle_0 + \Delta \vec{\varepsilon} \cdot \langle \partial_{\vec{\varepsilon}} \phi_i (\vec \varepsilon_0) \rangle_0 = \phi_i^* $,
    where $ \phi_i^* $ are target, error-free quantities defined as the central values of the observables $ \langle \phi_i(\vec \varepsilon) \rangle $ measured on the QCD+QED ensemble. A similar renormalization condition is applied to the scale-setting observable $t_0$. 
    This approach transfers all uncertainties to the target observable. On the dynamical QCD+QED side, propagating the uncertainties of $\phi_i$ to $ a_\mu^\hvp $ requires computing $\tfrac{d}{d{\phi_i}} a_\mu^\hvp = \tfrac{\partial}{\partial {\vec{\varepsilon}}} a_\mu^\hvp \cdot (\tfrac{\partial}{\partial {\vec{\varepsilon}}} \phi_i)^{-1}$, which can be computed directly on \texttt{A380a07b324} or approximated using the derivatives computed on the isoQCD ensemble: this should only introduce $\mathcal{O}((e^2+\Delta m_f)^2)$ IB corrections.
\end{enumerate}

In this proceeding, we adopt the first procedure, using error-free bare parameters shifts $\Delta \vec{\varepsilon}$ to compute observables in the isoQCD+RM123 framework. 
Moreover, we omit valence-disconnected Wick contractions and focus solely on valence-connected ones, while also neglecting IB effects from sea quarks (see Table~\ref{tab:diagrams}).

\section{Setup}

This section introduces the $C-$periodic boundary conditions used in our lattice QCD+QED simulations and outlines the ensembles and renormalization scheme employed in the analysis.

\subsection[C-periodic boundary conditions]{$\boldsymbol C-$periodic (or C$^{\star}$) Boundary Conditions}

Periodic Boundary Conditions (BCs) in finite-volume lattice simulations prevent the propagation of electrically charged states. To overcome this limitation, $C$-periodic (for brevity, $C^\star$ in the following) BCs were introduced~\cite{Kronfeld:1992ae,Polley:1993bn}. These boundary conditions allow charged hadrons to propagate in a finite lattice while preserving locality, gauge invariance, and translational invariance~\cite{Lucini:2015hfa}. $C^\star$ BCs are imposed by requiring that the gauge field transforms across the boundaries as $U_\mu(x + L_i \hat{i}) = U^*_\mu(x)$, while the fermion fields transform as $\psi(x + L_i \hat{i}) = C^{-1} \bar{\psi}^T(x)$ and $\overline{\psi}(x + L_i \hat{i}) = -\psi^T(x) C$, with $C$ being the charge conjugation matrix and $\hat{i}$ a unit vector in the $i=1,2,3$ spatial directions.

Finite-volume corrections to charged hadron masses exhibit power-law scaling under $C^\star$ BCs. Structure-dependent corrections appear only at $\mathcal{O}(1/L^4)$, in contrast to $\mathcal{O}(1/L^3)$ in other formulations like QED$_L$, significantly reducing finite-volume effects. $C^\star$ BCs also reduce finite-volume effects to $a_\mu^\hvp$, with respect to periodic BCs~\cite{Martins:2022hqb}.

A consequence of imposing $C^\star$ BCs is a weak violation of flavor conservation, as flavor-charged particles traveling around the torus transform into their antiparticles. However, this effect is exponentially suppressed with volume and is negligible for practical purposes in numerical simulations~\cite{Lucini:2015hfa}.

\subsection{Ensembles}
\label{sec:ensembles}
Table~\ref{tab:ensembles} lists the simulation parameters $\vec \varepsilon_0$ and $\vec \varepsilon_{\text{A380}}$ for the isoQCD (\texttt{A400a00b324}) and QCD+QED (\texttt{A380a07b324}) ensembles, while Table~\ref{tab:reno_scheme} defines their renormalization scheme.

In the isoQCD setup, where $m_u = m_d = m_s$, there are only two distinct light meson masses: $M_{\pi^\pm} = M_{K^\pm} = M_{K^0} = 398.5(4.7)\,\text{MeV}$ and $M_{D^\pm} = M_{D_s^\pm} = M_{D^0} = 1912.7(5.7)\,\text{MeV}$. This symmetry significantly reduces the number of meson mass derivatives required for the RM123 method. For comparison, the QCD+QED ensemble yields $M_{\pi^\pm} = M_{K^\pm} = 383.6(4.4)\,\text{MeV}$, $M_{K^0} = 390.7(3.7)\,\text{MeV}$, $M_{D^\pm} = M_{D_s^\pm} = 1926.4(7.8)\,\text{MeV}$, and $M_{D^0} = 1921.1(7.6)\,\text{MeV}$, as previously computed in Ref.~\cite{RCstar:2022yjz}.

\begin{table}[h]
    \centering
    \resizebox{.97\linewidth}{!}{
    \renewcommand{\arraystretch}{1.1}
    \begin{tabular}{|l|c|c|c|c|c|c|}
     \hline
     \multicolumn{1}{|c|}{Ensemble} & lattice   & $\beta$     & $\alpha$    & $\kappa_u$    & $\kappa_d = \kappa_s$   & $\kappa_c$  \\
     \hline\hline
     A400a00b324 & 64 $\times$ 32$^3$ & 3.24     & 0  & 0.13440733   & 0.13440733    & 0.12784    \\
     A380a07b324 & 64 $\times$ 32$^3$ & 3.24     & 0.007299  & 0.13459164   & 0.13444333    & 0.12806355 \\
     \hline
     \noalign{\vskip\doublerulesep\vskip-\arrayrulewidth} \cline{3-7}
     \multicolumn{2}{c|}{} & $ \Delta \beta$    & $ \Delta\alpha$ & $ \Delta m_u$ & $ \Delta m_d =  \Delta m_s$ & $ \Delta m_c$   \\
     \cline{3-7}
     \noalign{\vskip\doublerulesep\vskip-\arrayrulewidth} \cline{3-7}
     \multicolumn{2}{r|}{}   & 0     & 0.007299    & -0.00509422     & -0.000996117  & -0.00682735     \\
     \cline{3-7}
    \end{tabular}
    }
    \caption{Parameters of the isoQCD (first row) and  QCD+QED (second row) ensembles. The isoQCD ensemble has $\kappa_u = \kappa_d = \kappa_s$, while the QCD+QED ensemble has $\kappa_u > \kappa_d = \kappa_s$ to account for $m_u < m_d$, with degenerate down and strange quarks. The parameter shifts $\Delta \varepsilon_k$ for $\beta$, $\alpha=4\pi e^2$, and quark masses are listed at the bottom of the table. These shifts can be extracted directly since both lattices use the same simulation code and action. 
    }
    \label{tab:ensembles}
\end{table}

\begin{table}[h]
    \centering
    \renewcommand{\arraystretch}{1.1}
    \resizebox{.97\linewidth}{!}{
    \begin{tabular}{|l|l|l|l|l|l|}
     \hline
     \multirow{2}{*}{Observable}  & Physical & \multicolumn{2}{c|}{\rcstar target value} & \multicolumn{2}{c|}{Measured Values} \\
     \cline{3-6}
   & value    & {\footnotesize isoQCD}     & {\footnotesize QCD+QED}    & {\footnotesize isoQCD} & {\footnotesize QCD+QED} \\
     \hline\hline
     $\phi_0 = 8t_0(m_{K^\pm}^2 - m_{\pi^{\pm}}^2)$  & 0.992    & 0    & 0  & ---     & ---     \\
     $\phi_1 = 8t_0(m_{K^{\pm}}^2 + m_{\pi^{\pm}}^2 + m_{K^{0}}^2)$ & 2.26     & 2.11    & 2.11   & 2.107(50)   & 1.977(37)   \\
     $\phi_2 = 8t_0(m_{K^{0}}^2 - m_{K^\pm}^2)/\alpha_R$     & 2.36     & 0 & 2.36      & ---     & 3.39(14)    \\
     $\phi_3 = \sqrt{8t_0}(m_{D_s^\pm} + m_{D^0} + m_{D^\pm})$  & 12.0     & 12.1    & 12.1   & 12.068(36)  & 12.132(48)  \\
     \hline
     $\sqrt{8 t_0}$ / fm   & 0.415    & 0.415   & 0.415  &  &  \\
     $\alpha_R$    & 0.007297 & 0    & $\alpha^{\text{phys}}$    &  &  \\ \hline
    \end{tabular}
    }
    \caption{Renormalization scheme for isoQCD (\texttt{A400a00b324}) and QCD+QED (\texttt{A380a07b324}) ensembles, generated in~\cite{RCstar:2022yjz}. The scale is set using $\sqrt{8 t_0} = 0.415~\text{fm}$ from Ref.~\cite{Bruno:2016plf}, giving lattice spacings $a = 0.05393(24)~\text{fm}$ (isoQCD) and $a = 0.05323(28)~\text{fm}$ (QCD+QED). Physical values of $\phi_{0,1,2,3}$ are compiled using the experimental masses in~\cite{ParticleDataGroup:2024cfk}, without errors. 
    At leading order in ChPT, $\phi_i$ depend on quark mass combinations: $\phi_0 \propto (m_s - m_d)$, $\phi_1 \propto (m_u + m_d + m_s)$, $\phi_2 \propto (m_u - m_d)$, and $\phi_3 \propto m_c$.}
    \label{tab:reno_scheme}
\end{table}

\subsection{RM123: Feynman diagrams with our action and vector currents}
For this analysis, we use both local-local and conserved-local implementations of the vector current correlator $G(t)$, defined in Section~\ref{sec:hvp}. As a result, the leading IB effects arise from two sources: the action and the conserved current $V^c_\mu$ at the sink, where the action 
is described in detail in the \texttt{openQCD} and \texttt{openQxD}  documentations~\cite{openQCD:LuscherGauge, openQCD:LuscherDirac, openQxD:GaugeAction, openQxD:Dirac}. 
Table~\ref{tab:diagrams} provides a summary of all Feynman diagrams at first order in $\Delta m_f$ and $e^2$ which are required to compute the IB corrections to $G(t)$.
The diagrams are categorized by valence quark connections (connected or disconnected) and by the placement of IB insertions on valence, sea, or mixed quarks. 
Note that for $m_d = m_s$, the ``U-isovector'' current $\bar \psi_d \gamma_\mu \psi_d - \bar \psi_s \gamma_\mu \psi_s$ requires no valence-disconnected diagrams, as these cancel by $d$-$s$ symmetry.  This quantity will be the focus of future work.

\begin{table}[t]
    \footnotesize
    \renewcommand{\arraystretch}{1.1}
    \centering
    \resizebox{\linewidth}{!}{
    \begin{tabular}{|>{\centering\arraybackslash}m{1em}|>{\centering\arraybackslash}m{3em}|>{\centering\arraybackslash}m{5.5em}|>{\centering\arraybackslash}m{18em}|>{\centering\arraybackslash}m{12em}|}
    \cline{3-5}
    \multicolumn{2}{c|}{}  & \multicolumn{2}{c|}{from action}  & from $V^c_\mu$ at sink
    \\  \cline{2-5}
    \multicolumn{1}{c|}{}  & IB type     & mass    & QED & QED
    \\ \hline
    \multirow{3}{*}{\rotatebox[origin=c]{90}{valence connected~~~}}
     & vv  
     & \includegraphics[scale=0.085]{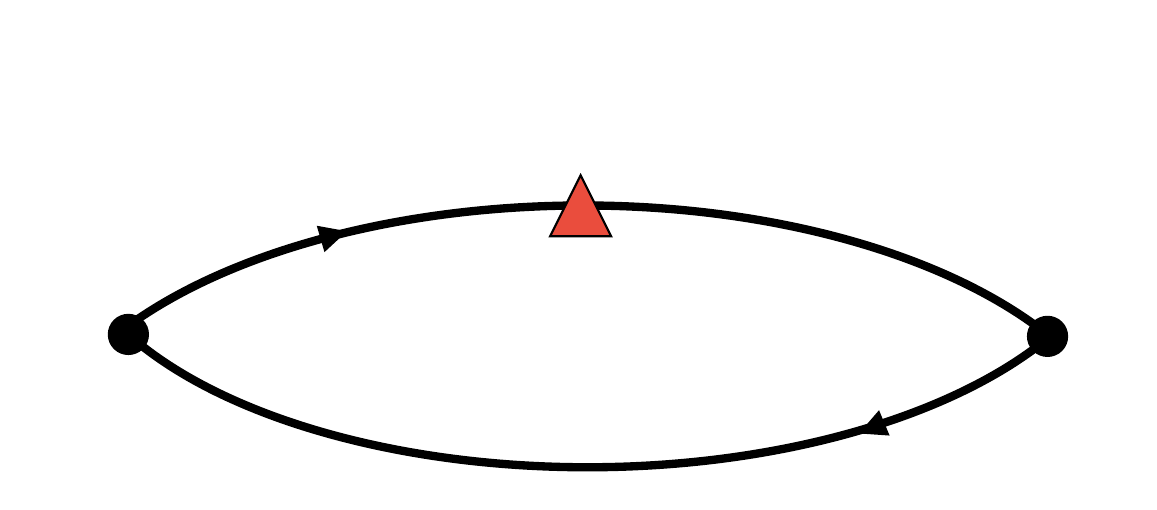} &
    \includegraphics[scale=0.085]{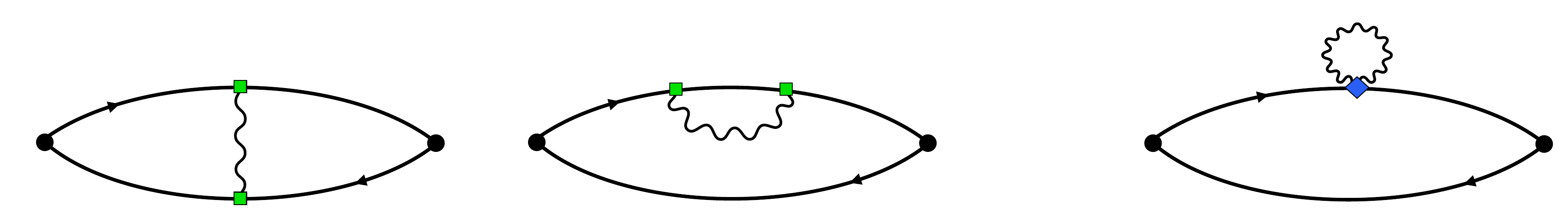} &
    \includegraphics[scale=0.085]{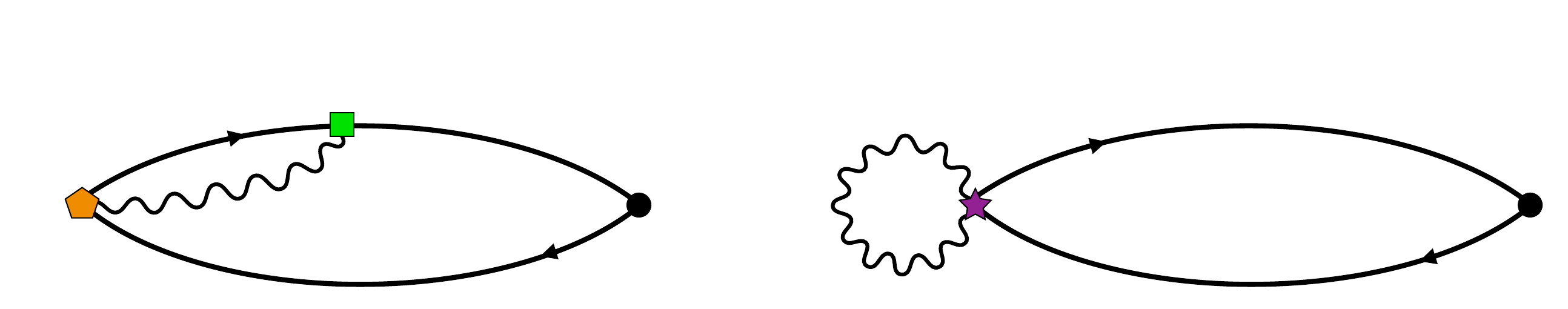}
    \\ \cline{2-5}
     & vs  
     & \includegraphics[scale=0.085]{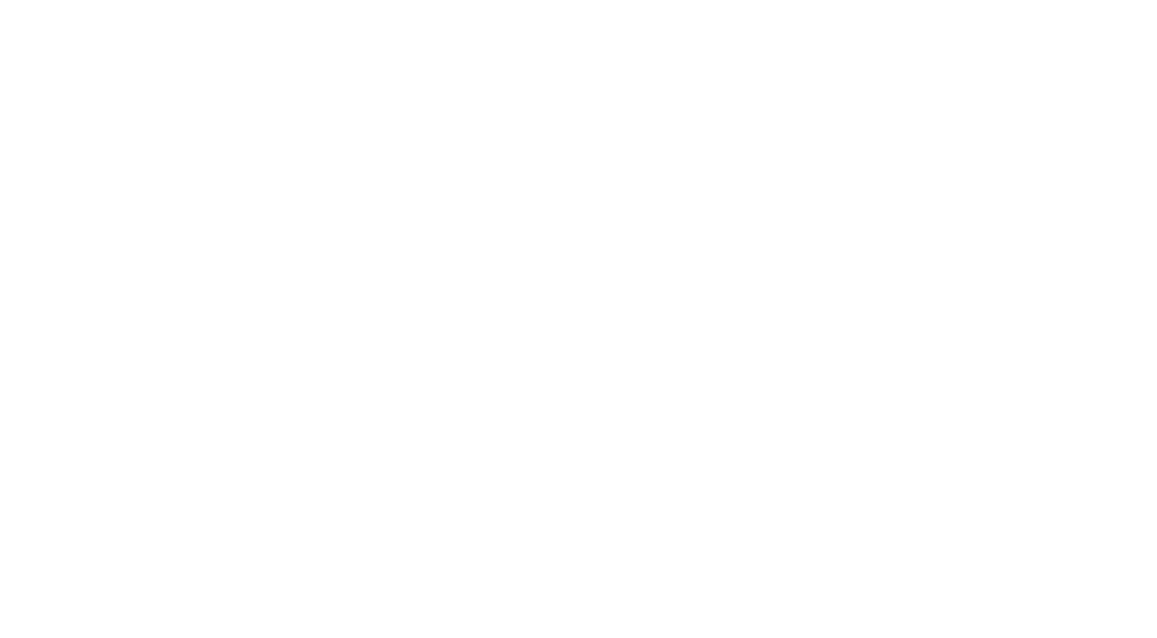} &
    \includegraphics[scale=0.085]{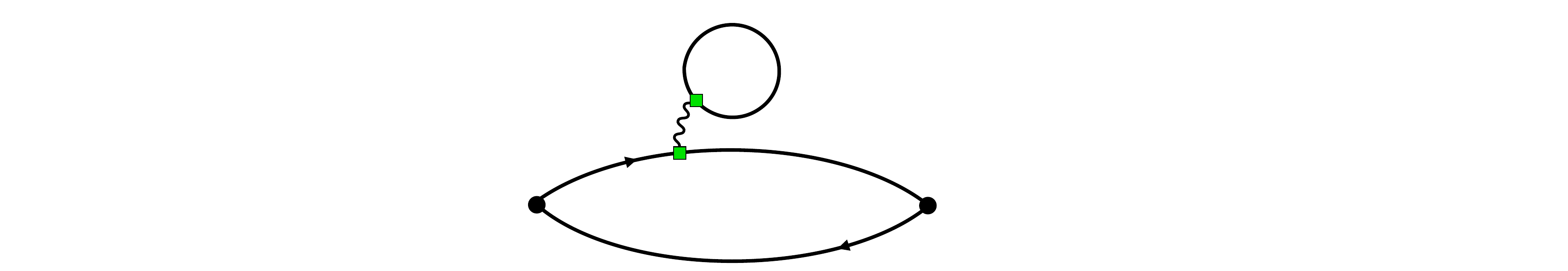} &
    \includegraphics[scale=0.085]{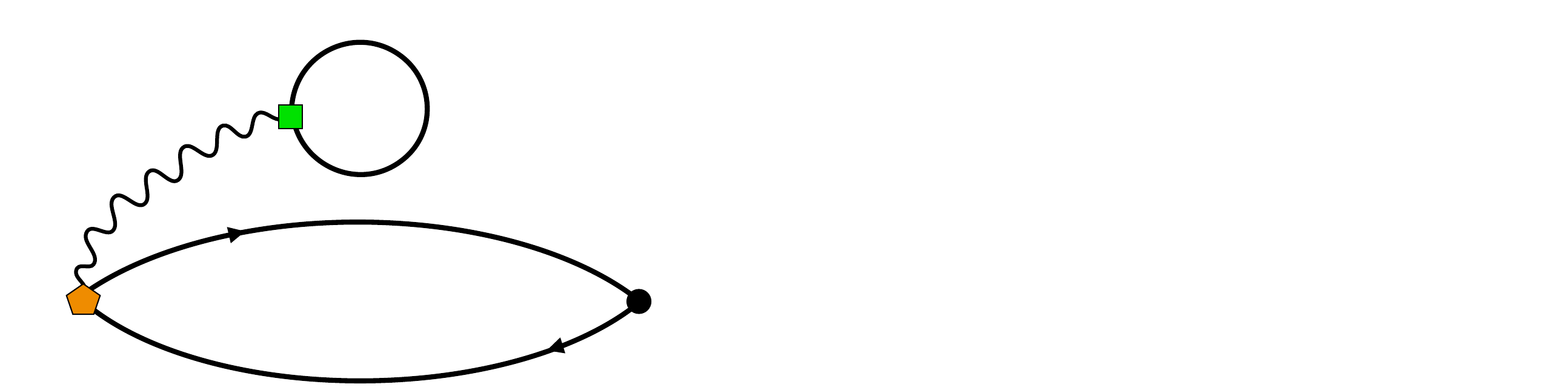}
    \\ \cline{2-5}
     & ss  
     & \includegraphics[scale=0.085]{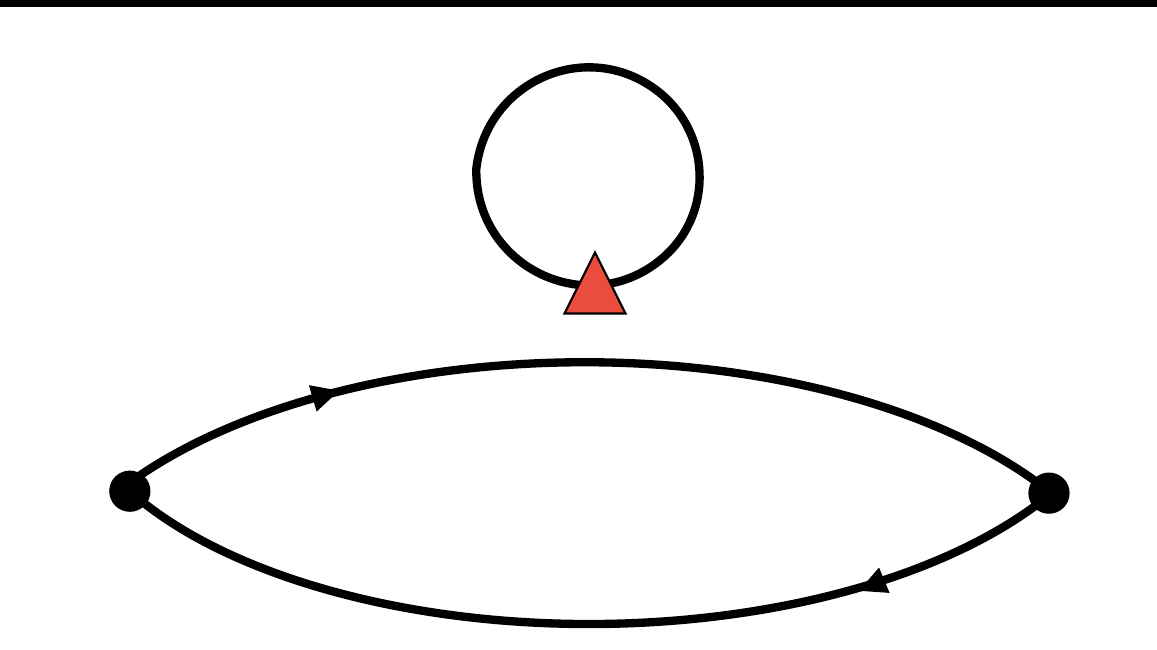} &
    \includegraphics[scale=0.085]{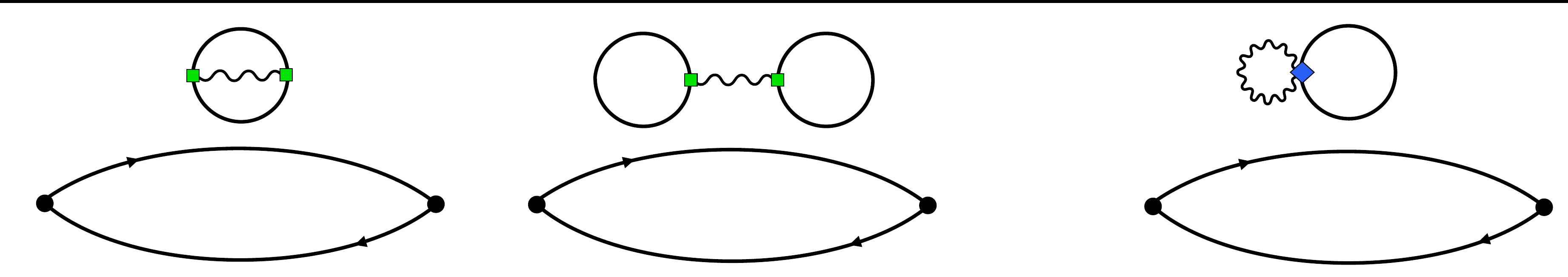} &
    \includegraphics[scale=0.085]{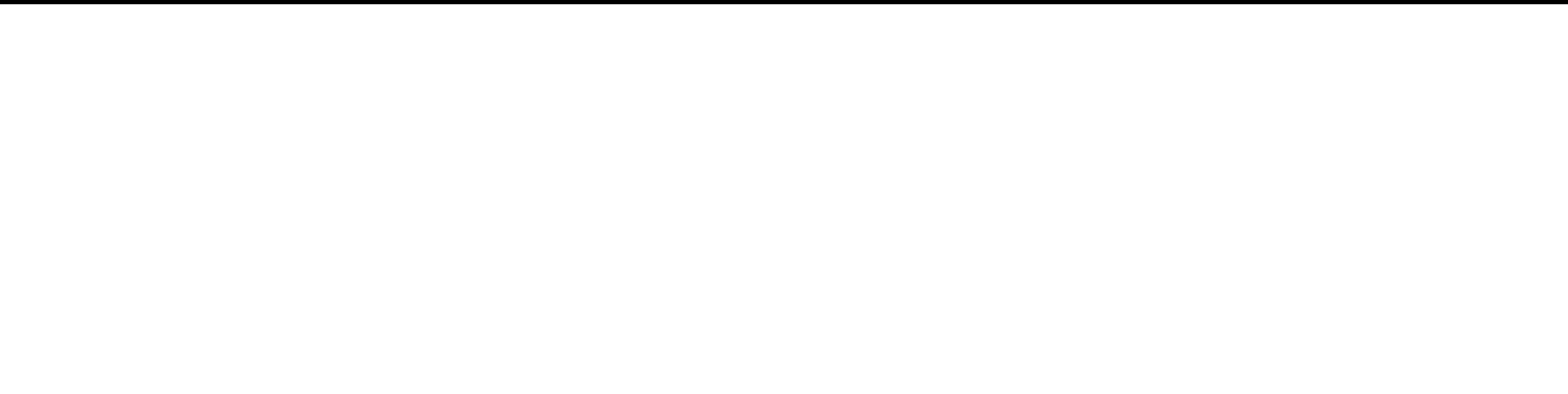}
    \\
    \hline
    \multirow{3}{*}{\rotatebox[origin=c]{90}{valence disc.ed~}}
     & vv   
     & \includegraphics[scale=0.085]{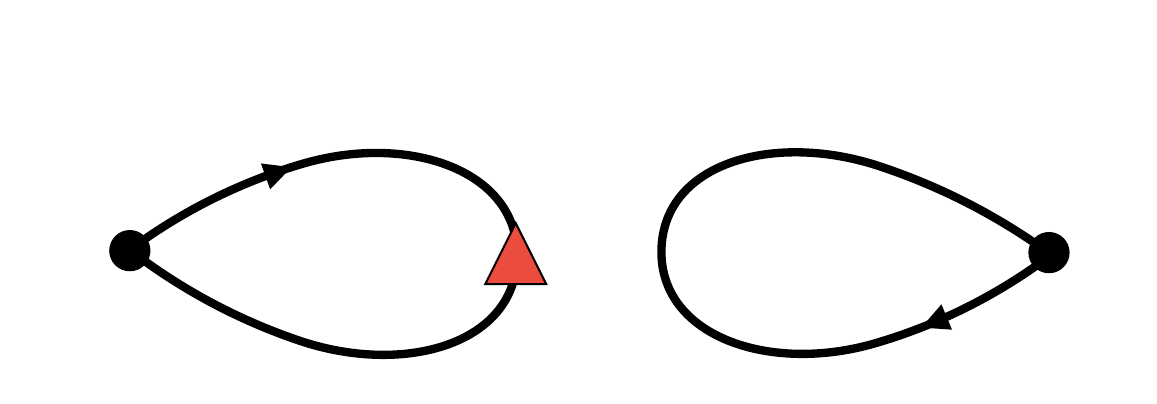} &
    \includegraphics[scale=0.085]{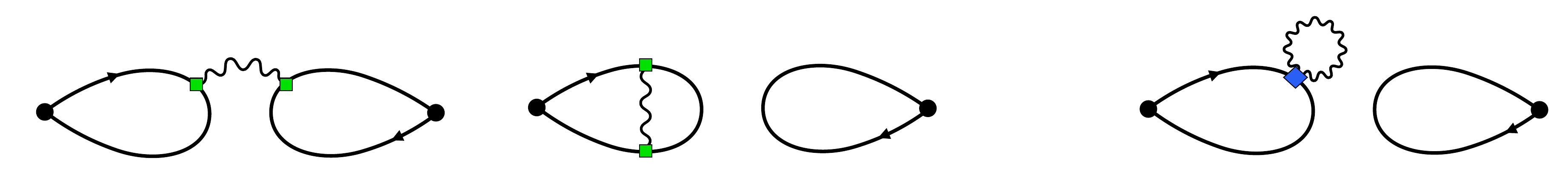} &
    \includegraphics[scale=0.085]{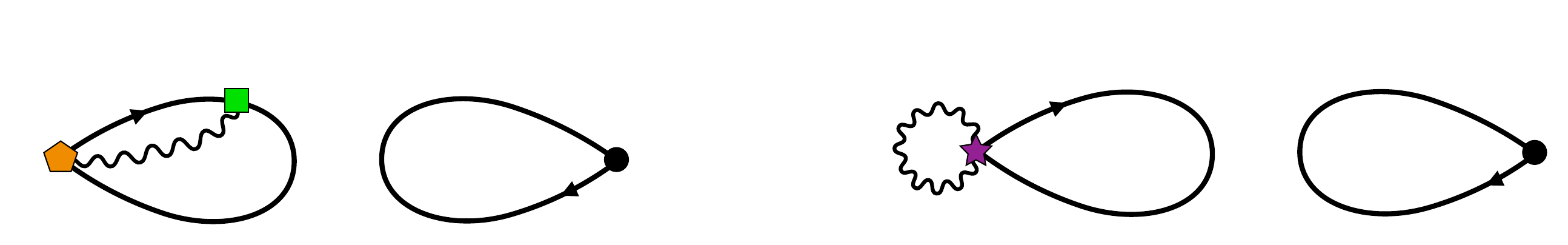}
    \\ \cline{2-5}
     & vs  
     & \includegraphics[scale=0.085]{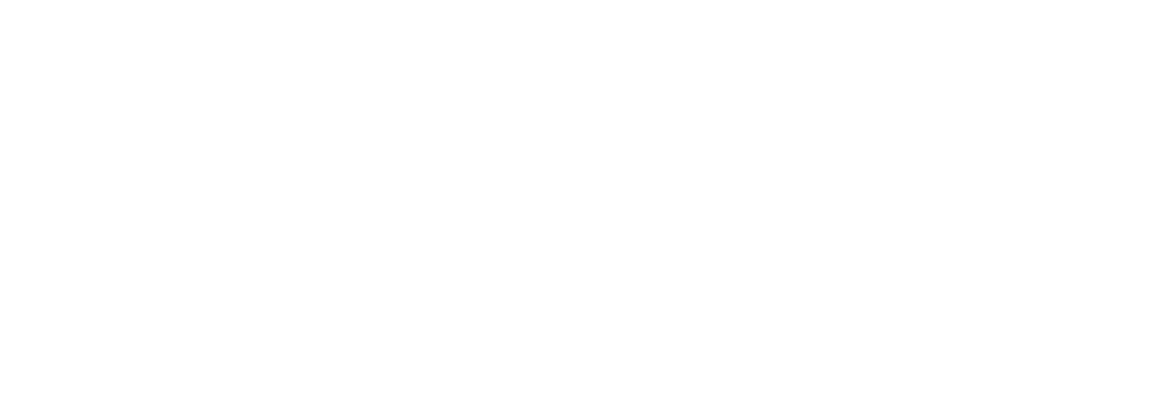} &
    \includegraphics[scale=0.085]{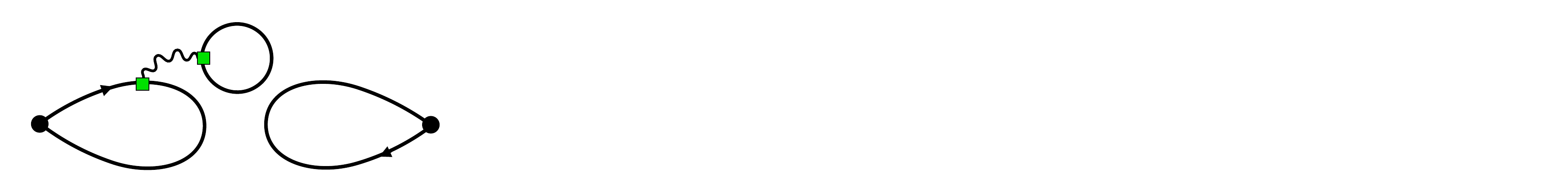} &
    \includegraphics[scale=0.085]{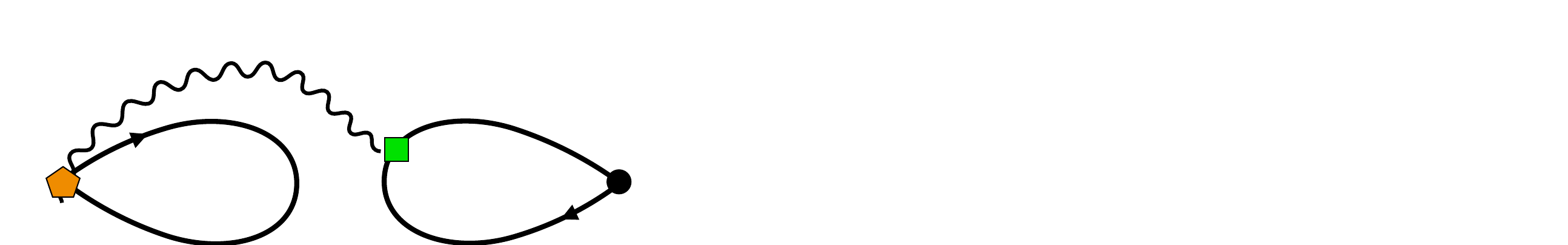}
    \\ \cline{2-5}
     & ss  
     & \includegraphics[scale=0.085]{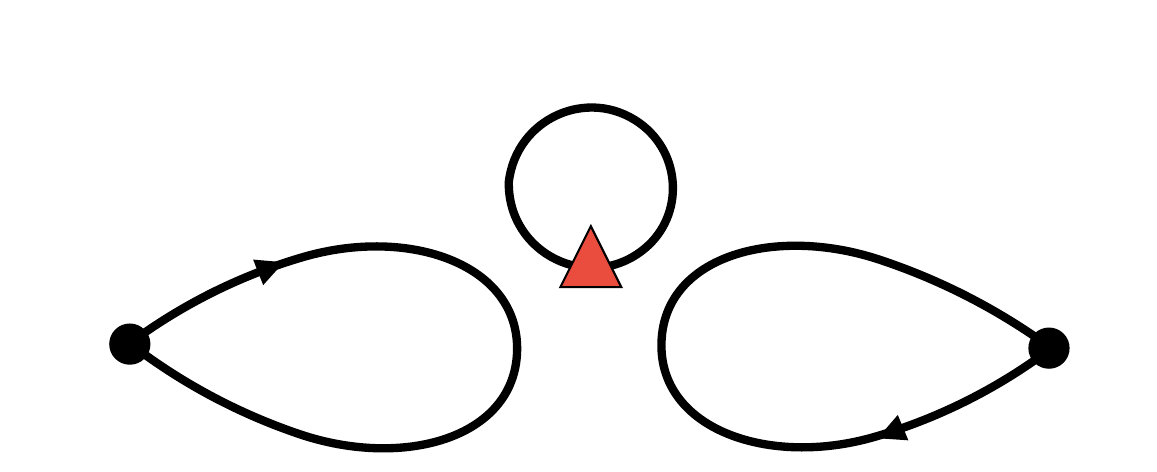} &
    \includegraphics[scale=0.085]{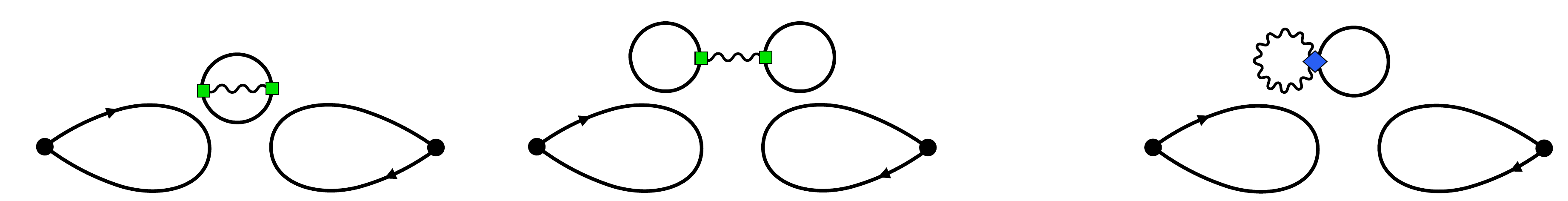} &
    \includegraphics[scale=0.085]{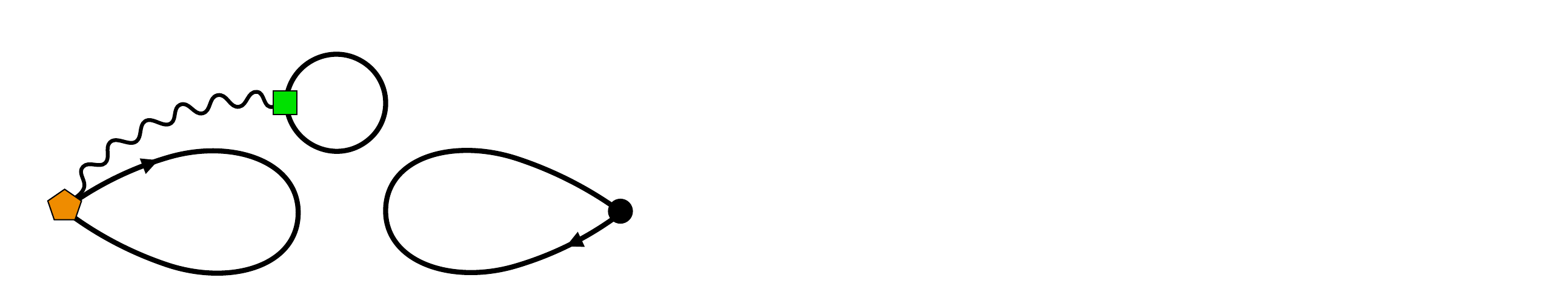}
    \\
    \hline
    \end{tabular}}
    \caption{
    Diagrams of IB contributions to a two-point function, in mass and QED sectors. Colored symbols denote operator insertions: red triangle (mass term, $\sum_x \bar{\psi}_f\psi_f$), green square (single photon insertion, from either $\partial_e D_W$ or $\partial_e D_{SW}$), 
    blue diamond (photon tadpole, from $\partial^2_e D_W$ only, because $\partial^2_e D_{SW}=0$). When using conserved-local correlator $G^{cl}(t)$, and thus conserved vector current $V^c_{k}$ at sink, additional insertions appear (last column): 
    orange pentagon (single photon, from $\partial_{e} V^c_k(0)$) and purple star (tadpole, from $\partial^2_{e} V^c_k(0)$). 
    Only vv-IB corrections to valence-connected diagrams (first row, top block) are included in this analysis, neglecting disconnected diagrams (bottom block) and sea quark contributions (vs, ss rows).
    }
    \label{tab:diagrams}
\end{table}

\section{Scheme matching and mass parameter shifts}
\label{sec:mass_shifts} 
The Eqs.~\eqref{eq:first}-\eqref{eq:last} provide the first-order expansions (denoted by superscript~$^{(1)}$) in the bare parameters $e^2$ and $\Delta m_f$ for the scheme-defining observables $\phi_i$ defined in Table~\ref{tab:reno_scheme}:
\begin{subequations}
    \label{eq:phis}
    \begin{align}
    &\phi_0^{(1)} = \phi_0^{(0)} = 0\,, \label{eq:first} \\
    &\phi_1^{(1)} = \phi_1^{(0)} + 16t_0^{(0)} m_{\pi^{\pm}}^{(0)} \Big[ 
    \Big(\sum\nolimits_{f=d,s} \Delta m_f (\partial_{m_f}m_{K^0})^{(0)} + e^2 (\partial_{e^2}m_{K^0})^{(0)}\Big) \nonumber\\*
    &\hspace{102pt} + 2\Big(\sum\nolimits_{f=u,d} \Delta m_f (\partial_{m_f}m_{\pi^\pm})^{(0)} + e^2 (\partial_{e^2}m_{\pi^\pm})^{(0)}\Big)\Big]\,,
    \\
    &\phi_2^{(1)} = \phi_2^{(0)} + 16t_0^{(0)} \frac{m_{K^0}^{(0)}}{4\pi e^2} \Big[
    \Big(\sum\nolimits_{f=d,s} \Delta m_f (\partial_{m_f}m_{K^0})^{(0)} + e^2 (\partial_{e^2}m_{K^0})^{(0)}\Big) \nonumber\\*
    &\hspace{102pt} - \Big(\sum\nolimits_{f=u,s} \Delta m_f (\partial_{m_f}m_{K^\pm})^{(0)} + e^2 (\partial_{e^2}m_{K^\pm})^{(0)}\Big)\Big]\,,
    \\
    &\phi_3^{(1)} = \phi_3^{(0)} + \sqrt{8t_0^{(0)}} \Big[
    \Big(\sum\nolimits_{f=u,c} \Delta m_f (\partial_{m_f}m_{D^0})^{(0)} + e^2 (\partial_{e^2}m_{D^0})^{(0)}\Big) \nonumber\\*
    &\hspace{102pt} + 2\Big(\sum\nolimits_{f=d,c} \Delta m_f (\partial_{m_f}m_{D^\pm})^{(0)} + e^2 (\partial_{e^2}m_{D^\pm})^{(0)}\Big)\Big]\,, \label{eq:last}
    \end{align}
\end{subequations}
where $X^{(0)}$ indicates that the observable $X$ is defined and computed in isoQCD.
The derivatives of scale $t_0$ are currently ignored and will be included alongside the IB effects from sea quarks. For the same reason, the additional equation for the lattice spacing, $a = a^{(0)} - \frac{1}{2} (\Delta \hat{t}_0 / \hat{t}_0)\,a^{(0)}$, where $\Delta \hat{t}_0 = \sum\nolimits_{f} \Delta m_f (\partial_{m_f}\hat{t}_0)^{(0)} + e^2 (\partial_{e^2}\hat{t}_0)^{(0)} 
+ \Delta \beta (\partial_\beta\hat{t}_0)^{(0)} = 0$, is also neglected.
Moreover, in our isoQCD setup, several mass degeneracies occur, due to $m_u = m_d = m_s$; in particular $K^{\pm}$ and $\pi^{\pm}$ are effectively the same particle, similarly for $D_s^{\pm} = D^\pm$. These degeneracies have been used to simplify several terms in Eqs.~\eqref{eq:first}-\eqref{eq:last}.

If we were to pursue strategy (b) described in Section~\ref{sec:intro}, we would need to fix the left-hand side of Eqs.~\eqref{eq:first}-\eqref{eq:last} to the target QED+QCD scheme and compute the mass shifts $\Delta m_f$ that satisfy the system of equations. Since we are here following strategy (a), we fix $\Delta m_f$ to the values given in Table~\ref{tab:ensembles} and compute the set of $\phi_i$ for isoQCD+RM123 with their errors:
\begin{equation}
    \phi_1 = 2.164(23)(4) \,,\quad
    \phi_2 = 3.602(47)(27) \,,\quad
    \phi_3 = 12.095(30)(1) \,,
\end{equation}
where the first and second brackets correspond to the statistical and systematic errors, respectively. These values can be compared to those computed directly on the {\texttt{A380a07b324}} ensemble.

\section{$\boldsymbol{a_\mu^\hvp}$ in QCD+QED}
\label{sec:hvp}

The HVP contribution to the $(g-2)$ of the muon, $ a_{\mu}^{\text{HVP}}$, is given by:
\begin{equation}
    a_{\mu}^{\text{HVP}} = \left( \frac{\alpha}{\pi} \right)^2 \sum_{f_1,f_2} \int_0^\infty dt \, G^{R,{ll}}_{f_1f_2}(t) {K}(t; m_{\mu}),
\end{equation}
where $ G_{f_1f_2}(t) = q_{f_1}q_{f_2}\,\frac{1}{3} \sum_{k=1}^3 \sum_{\vec{x}} \langle V^{f_1}_{k}(x) V^{f_2}_{k}(0) \rangle $, with $V^{f_1}_{k}(x)$ the vector current, and $ K(t; m_{\mu}) $ defined as in Ref.~\cite[Eq.~44]{DellaMorte:2017dyu}. Here, $ G^{R,ll}(t) $ represents the renormalized local-local correlator 
$G^{R,ll}(t) = Z_V G^{ll}(t) Z_V^T$, 
where $ G $ and $ Z_V $ are $ 4 \times 4 $ matrices in the flavor basis (see \cite{Gerardin:2019rua} for the renormalization pattern with $N_f=3$, $O(a)$-improved Wilson fermions).
The computation of $Z_V$ is briefly discussed in Sec.~\ref{sec:reno_const_LO}. 

Alternatively, we use the renormalized conserved-local correlator $G^{R,cl}(t) =  Z_V G^{cl}(t)$.

The IB corrections to $ a_{\mu}^{\text{HVP}} $ are categorized into two main types:
\begin{enumerate}\itemsep0pt
    \item  Corrections to correlators:
     \begin{align}
   \delta_{G}a_{\mu}^{\text{HVP}} & = \left( \frac{\alpha}{\pi} \right)^2 \int dt\, Z_V^{(0)}  \Delta G^{ll}(t) Z_V^{(0)^T}\, {K}(t; m_{\mu}) \label{eq:delta_G_correction}   \\
   \Delta G^{ll}(t)  & = \sum_f  \Delta m_f \left(\frac{\partial G^{ll}(t)}{\partial m_f} \right)^{(0)} + e^2 \left( \frac{\partial G^{ll}(t)}{\partial e^2} \right)^{(0)}
   \label{eq:delta_G}
  \end{align}%
    \item Corrections to renormalization constants:
   \begin{align}
   \delta_{Z} a_{\mu}^{\text{HVP}} & = \left( \frac{\alpha}{\pi} \right)^2 \int dt\, \left[Z_V^{(0)} G^{ll}(t)^{(0)} \Delta Z_V^T +  \Delta Z_V G^{ll}(t)^{(0)} Z_V^{(0)^T}\right]\, {K}(t; m_{\mu}) \label{eq:delta_Z_correction} \\
   \Delta Z_V  & = \sum_f  \Delta m_f \left(\frac{\partial Z_V}{\partial m_f}\right)^{(0)}+e^2 \left(\frac{\partial Z_V}{\partial e^2}\right)^{(0)}%
   \label{eq:delta_Z}%
   \end{align}%
\end{enumerate}%

If sea-sea effects were included, the IB corrections to the lattice spacing would also require considering the derivative of the kernel with respect to $ a $, due to its implicit dependence $ t = a \hat{t} $.

\subsection{Renormalization constants $\boldsymbol{Z_V}$ at leading order}
\label{sec:reno_const_LO}
The renormalization conditions are defined in the adjoint basis of SU(4) generators $\lambda_{3}, \lambda_8, \lambda_{15}$, and the identity $\lambda_0 = \mathbb{1}$ as outlined in Refs.~\cite{Bhattacharya:2005rb, Gerardin:2019rua}:
\begin{equation}
        V_{\mu}^{em} = \sum_{f=u,d,s,c} Q_f \bar\psi_f \gamma_\mu \psi_f = \frac{1}{3} V_{\mu}^{0} + V_{\mu}^{3} + \frac{1}{\sqrt{3}} V_{\mu}^{8} - \frac{1}{\sqrt{6}} V_{\mu}^{15}
\end{equation}
where the adjoint currents are $V_\mu^{0,3,8} = \frac{1}{2} \text{tr}(\lambda_{0,3,8} \mathcal V)$ and $V_\mu^{15} = \text{tr}(\lambda_{15} \mathcal V)$, with $[\mathcal{V}]_{f_1 f_2} = \bar\psi_{f_1} \gamma_\mu \psi_{f_2}$. In the adjoint basis, we define $\tilde Z_V$ as follows:

\begin{equation}
    \left[ \tilde{Z}_V \right]_{ab} = \lim_{x_0 \to \infty} \tilde{G}^{cl}_{ad} \cdot (\tilde{G}^{ll})^{-1}_{db}, \quad a, b = 0, 3, 8, 15.
    \label{eq:zv_def}
\end{equation}

In this work, we are neglecting the disconnected terms. As a consequence, the renormalization constant matrices, when brought back in the flavor bases, are diagonal and computed to be:
\begin{align}
    {Z}^{A400}_V &= \text{diag}(
    0.6771(3),\, 0.6771(3),\, 0.6771(3),\, 0.6050(8)
    )\,,
    \\
    {Z}^{A380}_V &= \text{diag}(
    0.6775(6),\, 0.6793(7),\, 0.6793(7),\, 0.6048(9)
    )\,.
\end{align}

\subsection{Leading-order results for $\boldsymbol{a_\mu^\hvp}$ from connected correlators}
\label{sec:lo-hvp}

Results for $a_\mu^{\text{HVP}}$ from leading-order (LO) connected correlators are given in columns 1 and 3 of Table~\ref{tab:results_amu} (Conclusions) for ensembles {\tt A400a00b324} and {\tt A380a07b324}. Here ``leading-order'' is used to indicate 2-point functions only, rather than QCD only, as QED effects are inherently included in the QCD+QED ensemble {\tt A380a07b324}.

All calculations were performed using 2000 configurations with 4 point sources per configuration, neglecting disconnected contributions. At large Euclidean times, the correlator tails were reconstructed using single-exponential fits with $t_{\text{cut}} \in (1.2, 1.3) \,\text{fm}$.

\subsection{Corrections from derivatives of correlator}
The derivatives ${\partial_{m_f}}\,G(t)$ and ${\partial_e^2}\,G(t)$ correspond, diagrammatically, to the first row of Table~\ref{tab:diagrams}, neglecting disconnected and sea effects.
The $\mathcal{O}(e^2)$ insertions come from derivatives of the Wilson-Dirac operator $D_{\text{W}}$  and the Sheikholeslami–Wohlert (SW) term $\delta D_{\text{SW}}$ in the Dirac operator~\cite[Eqs.~8,~10,~13]{openQxD:Dirac}, leading to a total of 8 Wick contractions (11 for conserved-local) for each flavor.

The corrections to the tail parameters \(A\) and \(m_{\text{eff}}\) are defined as \(A = A^{(0)} + \Delta A\) and \(m_{\text{eff}} = m_{\text{eff}}^{(0)} + \Delta m_{\text{eff}}\). The parameters $\Delta A$ and $\Delta m_{\text{eff}}$ are extracted from a two-parameter linear fit:
\begin{equation}
    \frac{G^{(1)}(x_0) - G^{(0)}(x_0)}{G^{(0)}(x_0)} = \frac{\Delta A}{A^{(0)}} - x_0 \Delta m_{\text{eff}}.
\end{equation}
To estimate systematic effects, this procedure is repeated over different fit ranges for the light quarks.

\subsection{Corrections from derivatives of $\boldsymbol{Z_V}$}
The correction to $ Z_V $ are defined in Eq.~\eqref{eq:delta_Z}.
We expect $ \delta_{Z_V} a_{\mu}^\hvp $ in the case of local-local discretization to be approximately twice as large as the corresponding correction in the conserved-local case, provided that $ {a_{\mu}^\hvp}$ from both $cl$ and $ll$ prescriptions agree at leading order.

The Wick contractions needed to compute $\partial_{\varepsilon_i} Z_V$ are the same as those for $\partial_{\varepsilon_i} G(x_0)$. These are fitted to the following expression, derived from Eq.~\eqref{eq:zv_def}:

\begin{equation}
    \frac{\partial Z_{V_R V_l}}{\partial \varepsilon_i} = \lim_{x_0 \to \infty} 
    \left[
    \frac{\partial G^{cl}}{\partial \varepsilon_i}(x_0) - G^{cl}(x_0) 
    \left( G^{ll}(x_0) \right)^{-1} 
    \frac{\partial G^{ll}}{\partial \varepsilon_i}(x_0)
    \right] \cdot \left( G^{ll}(x_0) \right)^{-1}.
\end{equation}

Tables \ref{tab:combined_corrections} and \ref{tab:results_amu} summarize IB corrections and total $a^\hvp_\mu$, respectively. Results for the total $a^\hvp_\mu$ are provided for the QCD+QED and isoQCD+RM123 setups, as well as for the isoQCD case.

\begin{table}[ht]
    \centering
    \renewcommand{\arraystretch}{1.15}
    \resizebox{\textwidth}{!}{
    \begin{tabular}{|l|l|l|l|l|}
     \hline
     \multicolumn{4}{|l}{{Corrections from renormalization constants} $\times 10^{10}$} & ~ \\ \hline
     \multirow{3}{*}{$ll$}
  & $ \delta_{Z_v} a_{\mu}^{uu} $
  & $ -510(33)  \Delta m_u -22(3) e^2 $
  & $2.60(17) - 2.02(28)$
  & $0.58(44)$  \\
  & $ \delta_{Z_v} a_{\mu}^{dd} $
  & $ 0.016(23)  \Delta m_u -128(8)  \Delta m_{d} -1.4(2) e^2 $
  & $0.127(8) - 0.128(18)$
  & $0.001(26)$    \\
  & $ \delta_{Z_v} a_{\mu}^{cc} $
  & $ 0.003(4)  \Delta m_u -6.37(2)  \Delta m_c -0.578(2) e^2$
  & $0.04347(14) - 0.05302(18)$
  & $-0.00954(34)$    \\ \hline
     \multirow{3}{*}{$cl$}
  & $ \delta_{Z_v} a_{\mu}^{uu} $
  & $ -252(16)  \Delta m_u -11(2) e^2$
  & $1.28(8) - 1.01(18)$
  & $0.27(26)$  \\
  & $ \delta_{Z_v} a_{\mu}^{dd} $
  & $ 0.008(11)  \Delta m_u -63(4)  \Delta m_{d} -0.68(11) e^2 $
  & $0.063(4) - 0.062(10)$
  & $0.000(14)$   \\
  & $ \delta_{Z_v} a_{\mu}^{cc} $
  & $ 0.0012(16)  \Delta m_u -2.516(9)  \Delta m_c -0.228(8) e^2$
  & $0.01717(6) - 0.0209(7)$
  & $-0.0038(1)$    \\ \hline
     \multicolumn{4}{|l}{{Corrections from correlator} $\times 10^{10}$}  & ~ \\ \hline
     \multirow{3}{*}{$ll$}
  & $ \delta_{G} a_{\mu}^{uu} $
  & $ -4364(266)  \Delta m_u -216(14) e^2$
  & $22.2(1.4) - 19.8(1.3)$
  & $2.4(2.6)$  \\
  & $ \delta_{G} a_{\mu}^{dd} $
  & $ -1091(67)  \Delta m_{d} -13.5(9) e^2 $
  & $1.09(7) - 1.24(8)$
  & $-0.15(15)$   \\
  & $ \delta_{G} a_{\mu}^{cc} $
  & $ -59.2(3)  \Delta m_c -3.119(13) e^2$
  & $0.404(20) - 0.2861(12)$
  & $0.118(22)$    \\ \hline
     \multirow{3}{*}{$cl$}
  & $ \delta_{G} a_{\mu}^{uu} $
  & $ -4591(288)  \Delta m_u -227(15) e^2$
  & $23.4(1.5) - 20.8(1.4)$
  & $2.6(2.8)$  \\
  & $ \delta_{G} a_{\mu}^{dd} $
  & $ -1148(72)  \Delta m_{d} -14.2(1.0) e^2$
  & $1.14(7) - 1.30(9)$
  & $-0.16(16)$   \\
  & $ \delta_{G} a_{\mu}^{cc} $
  & $ -57.2(2)  \Delta m_c -3.295(14) e^2$
  & $0.391(14) - 0.3022(13)$
  & $0.088(15)$   \\ \hline
    \end{tabular}}
    \caption{IB corrections from the renormalization constant $Z_V$ and the correlator $G$, for local-local ($ll$) and conserved-local ($cl$) currents. Due to $d\)-\(s$ flavor symmetry (\(\Delta m_d = \Delta m_s\)), $d$ can represent either strange or down quarks. 
    The last column sums the second-to-last column with fully correlated errors: these values are provided as reference, but are not directly added to isoQCD results in Table~\ref{tab:results_amu}; see its caption for details.}
    \label{tab:combined_corrections}
\end{table}

\begin{table}[h]
    \centering
    \renewcommand{\arraystretch}{1.15}
    \resizebox{.85\textwidth}{!}{
    \begin{tabular}{|c|c|c||c|c|c|c|}
    \cline{2-7}
    \multicolumn{1}{c|}{}    & \multicolumn{2}{c||}{isoQCD} & \multicolumn{2}{c|}{isoQCD+RM123} & \multicolumn{2}{c|}{QCD+QED}   \\
    \cline{2-7}
    \multicolumn{1}{c|}{}    & $ll$  & $cl$    & $ll$     & $cl$  & $ll$  & $cl$  \\
    \hline
    $a_\mu^{u} \times 10^{10}$   & 188.5(1.9)  & 186.5(2.0)    & 192.4(2.0)    & 189.2(2.0) & 194.0(2.3) & 192.2(2.2) \\
    $a_\mu^{d/s} \times 10^{10}$ & 47.1(5)     & 46.6(5)   & 47.0(5)    & 46.4(5)    & 47.2(6)    & 46.8(6)    \\
    $a_\mu^{c} \times 10^{10}$   & 7.59(3)     & 5.99(3)   & 7.73(3)    & 6.07(3)    & 7.55(4) & 5.95(4) \\
    \hline
    \end{tabular}
    }
    \caption{Results for $a_\mu^{\text{HVP}}$ in three setups: isoQCD (left) and two methods for including QED effects: isoQCD+RM123 (center) and dynamical QCD+QED (right). Results for isoQCD+RM123 are not obtained by simply adding IB corrections from Table~\ref{tab:combined_corrections} to the isoQCD results. Instead, consistent fit ranges and tail reconstructions are used across isoQCD and IB corrections, ensuring all correlations are properly handled.}
    \label{tab:results_amu}
\end{table}

\section{Outlook and Conclusions}
\label{sec:conclusions}

This work compares two methods for computing IB effects to the HVP, including all valence-connected terms. Sea-valence and sea-sea IB effects, yet to be added, are computed on the same ensemble \texttt{A400a00b324}. A discussion of these effects, restricted to the $N_f=3$ case, is contained in Ref.~\cite{Cotellucci}. Results from isoQCD+RM123 and dynamical QCD+QED (Table~\ref{tab:results_amu}) show slight ($\lesssim 1\sigma$) incompatibility, underscoring the importance of sea IB effects to achieve full consistency.

Future work will use the isovector current $\bar{d} \gamma_\mu d - \bar{s} \gamma_\mu s$, which is well-defined and free of disconnected diagrams. 
The initial analysis will focus on the intermediate time window and omit reconstruction of the long-distance piece.

This comparison is a first step towards a systematic comparison of the dynamical QCD+QED approach and perturbative treatment of IB corrections; a final answer will require a variety of observables and ensembles with smaller pion masses, larger volumes, and finer lattice spacings.

\section*{Acknowledgments}

We acknowledge access to Piz Daint at the Swiss National Supercomputing Centre, Switzerland under the ETHZ's share with the project IDs eth8 and s1196. The funding from the SNSF (Project No. 200021\_200866) is gratefully acknowledged. The authors gratefully acknowledge the computing time granted by the Resource Allocation Board and provided on the supercomputer Lise and Emmy at NHR@ZIB and NHR@Göttingen as part of the NHR infrastructure. The calculations for this research were partly conducted with computing resources under the project bep00085 and bep00102.
AC's research is funded by the Deutsche Forschungsgemeinschaft (DFG, German Research Foundation) - Projektnummer 417533893/GRK2575 ``Rethinking Quantum Field Theory''.
The work was supported by the Poznan Supercomputing and Networking Center (PSNC) through grant numbers 450 and 466. This research was supported in part by grant NSF PHY-2309135 to the Kavli Institute for Theoretical Physics (KITP).

\end{document}